\documentstyle[12pt,aasms4]{article}

\received{1997 November 10}
\accepted{1998 February 23}

\slugcomment{{\em The Astrophysical Journal}, in press} 

\lefthead{Montes et al.}
\righthead{Radio Observations of SN 1980K}

\def\Mdot{\dot{M}}
\def\mdot{\dot{M}}
\def\Msun{M_{\sun}}

\begin{document}

\title{Radio Observations of SN 1980K: Evidence for Rapid Presupernova
Evolution} 

\author{Marcos J. Montes\altaffilmark{1}}
\affil{Naval Research Laboratory, Code 7214, Washington, DC 20375-5320;
montes@rsd.NRL.Navy.mil}
\authoraddr{Code 7214, Washington, DC, 20375}

\author{Schuyler D. Van Dyk\altaffilmark{2}}
\affil{Dept. of Physics \& Astronomy, UCLA, Los Angeles, CA 90095;
vandyk@astro.ucla.edu}

\author{Kurt W. Weiler}
\affil{Naval Research Laboratory, Code 7214, Washington, DC 20375-5320;
weiler@rsd.NRL.Navy.mil}

\author{Richard A. Sramek}
\affil{National Radio Astronomy Observatory, P.O. Box 0, Socorro, NM
87801; dsramek@nrao.edu}

\and
\author{Nino Panagia\altaffilmark{2}}
\affil{Space Telescope Science Institute, 3700 San Martin Drive, Baltimore, 
MD 21218; panagia@stsci.edu}
\altaffiltext{1}{Naval Research Lab/National Research Council Cooperative
Research Associate}
\altaffiltext{2}{Visiting scientist.}
\altaffiltext{3}{Affiliated with the Astrophysics Division, Space
Science Department of ESA.}

\begin{abstract}
New observations of SN 1980K made with the VLA at 20 and 6 cm from 1994
April through 1996 October show that the supernova (SN) has undergone a
significant change in its radio emission evolution, dropping by a
factor of $\sim 2$ below the flux density $S\propto t^{-0.73}$
power-law decline with time $t$ observed earlier. However, although $S$
at all observed frequencies has decreased significantly, its current
spectral index of $\alpha= -0.42\pm0.15$ ($S \propto \nu^{+\alpha}$) is
consistent with the previous spectral index of
$\alpha=-0.60_{-0.07}^{+0.04}$.

It is suggested that this decrease in emission may be due to the SN
shock entering a new region of the circumstellar material which has a
lower density than that expected for a constant speed ($w$), constant
mass-loss rate ($\mdot$) wind from the progenitor.  If such an
interpretation is correct, the difference in wind and shock speeds
appears to indicate a significant evolution in the mass-loss history of
the SN progenitor $\sim 10^4$ years before explosion, with a
change in circumstellar density ($\propto \mdot/w$) occurring over a
time span of $\lesssim 4 $ kyr.  Such features could be explained in
terms of a fast ``blue-loop'' evolutionary phase of a relatively
massive pre-SN progenitor star.
If so, we may, for the first time, provide a stringent constraint on the mass
of the SN progenitor based solely on the SN's radio emission.
\end{abstract}


\section{Introduction}
SN 1980K [RA(1950.0)$=20^{\rm h}34^{\rm m}26\fs68 \pm 0\fs01;$
Dec(1950.0)$ = +59\arcdeg55\arcmin56\farcs5 \pm 0\farcs2$] in NGC 6946
was discovered in 1980 October by Wild \markcite{wild}(1980), and an
initial, unsuccessful attempt was made on 1980 November 3 to detect it
at 6 cm with the Very Large Array (VLA).\footnote{The VLA is operated
by the National Radio Astronomy Observatory (NRAO) of Associated
Universities, Inc., under a cooperative agreement with the National
Science Foundation.} However, only 35 days after optical maximum SN
1980K was detected at 6 cm on 1980 December 4 and has been regularly
monitored at 20, 6, and occasionally 2 cm since then. Results from 1980
November through 1984 August are presented in Weiler et
al.\ \markcite{w86}(1986), and results from 1984 November through 1990
December are presented in Weiler et al.\ \markcite{w92}(1992). By 1990
December the flux density had fallen to $S<0.3$ mJy at 20 and 6 cm,
which made it difficult to observe with the VLA using typical, short
``snapshot'' observations, and regular monitoring was terminated.

However, optical imaging and spectra taken after 1990 indicated that SN
1980K was still a detectable, albeit relatively constant optical
luminosity source, particularly at H$\alpha$ (Leibundgut
\markcite{l94}1994).  Late-time optical and radio emission are
predicted to be correlated (Chevalier \& Fransson \markcite{c94}1994),
and in several examples are indeed observed to be related (see, e.g.,
Montes et al.\ \markcite{m97}1997, Van Dyk et al.\ \markcite{v98}1998
and references therein).  Therefore, deeper and more sensitive VLA
observations were made in 1994, 1995, and 1996 at both 20 and 6 cm, to
attempt to obtain new measurements of SN 1980K below our previous
sensitivity limits.  Interestingly, these new data indicate that SN
1980K has, in fact, {\em not} continued along its previously observed
power-law decline, but has dropped sharply in radio flux density in the
interval between 1990 and 1994.  Additionally, recent measurements
(Fesen \markcite{f98}1998) indicate that SN~1980K has also faded in the
optical over the last several years, possibly starting with a 20\%
decline by 1994 as reported by Fesen, Hurford, \& Matonick
(\markcite{f95}1995).  The new radio measurements are presented here,
and they indicate a significant change in the evolution of the radio
emission of SN 1980K from earlier, model-based expectations.

\section{New Observations}
The new observations of SN 1980K were made with the VLA at 20 cm (1.46
GHz) and 6 cm (4.89 GHz) on  1994 April 10, 1995 June 22 (20 cm only),
and  1996 October 13--14, when the VLA was in the ``A'' configuration,
to reduce the confusion with background emission from the parent galaxy
NGC 6946.  Each observation was accompanied by measurement at each
frequency of a ``secondary,'' possibly variable, calibrator (2021+614),
with a well-known position, making it suitable as a phase reference. To
establish an absolute flux density scale, a ``primary'' calibrator (3C
286, except on 1995 June 22, when 3C 48 was used) was also observed and
used to calibrate the amplitude of 2021+614 for each observing session.
Additional observations of SN 1980K on 1994 April 01 and 1994 June 16
were kindly provided by C.~Lacey (\markcite{l97}1997; see also
Lacey, Duric, \& Goss 1997).

\subsection{Calibration}
The VLA is described in a number of publications (e.g., Thompson et
al.\ \markcite{t80}1980; Hjellming \& Bignell \markcite{h82}1982;
Napier, Thompson, \& Ekers \markcite{n83}1983); the general procedures
for RSN observations, with the sources of possible error, are discussed
in Weiler et al.\ \markcite{w86}(1986). The primary calibrators
employed here, 3C 48 for the 1995 June observation and 3C 286 for the
rest, were assumed to be constant in flux density with time and to have
the flux densities at 20 cm and 6 cm of $S_{\rm 20\ cm}({\rm
3C\ 48})=15.97$ Jy, $S_{\rm 20\ cm}({\rm 3C\ 286})=14.75$ Jy, and
$S_{\rm 6\ cm}({\rm 3C\ 286})=7.49$ Jy.

\begin{figure}
\plotfiddle{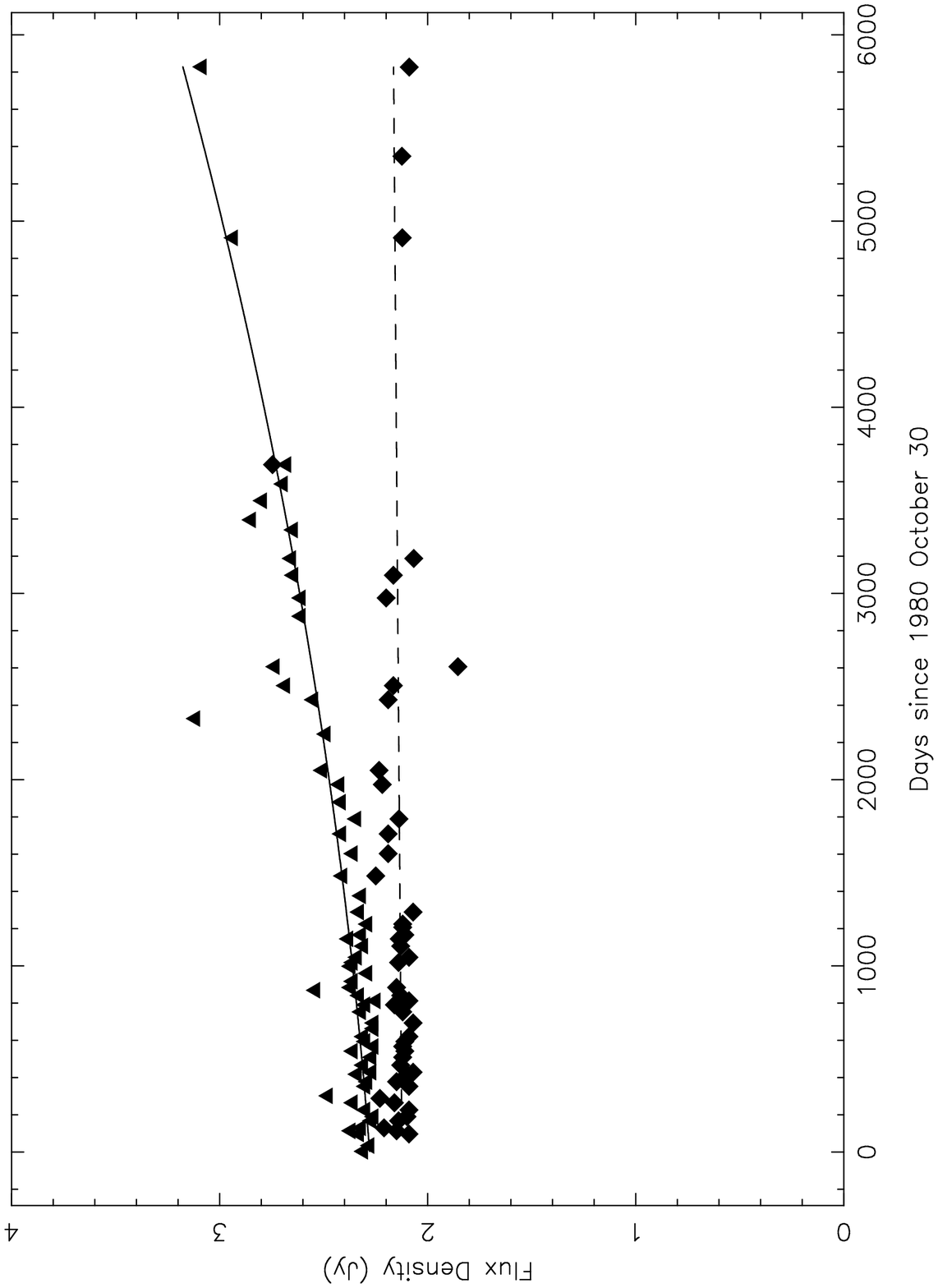}{5.5in}{270}{70}{70}{-250}{450}
\caption{
Flux density evolution of the secondary
calibrator, 2021+614, with 6 cm ({\em filled triangles}) and 20 cm
({\em filled diamonds}) data shown. Smoothed curves ({\em solid} for 6
cm and {\em dashed} for 20 cm) have been added to guide the eye; the
curves were not used in the calibration procedure.\label{fig:cal}}
\end{figure}

The secondary calibrator, 2021+614, was used as the phase (position)
[RA(1950.0)=$20^{\rm h} 21^{\rm m} 13\fs300$, Dec(1950.0)=$61\arcdeg
27\arcmin 18\farcs156$] and amplitude (flux density) reference for the
observations of SN 1980K.  Its flux density (measured against 3C 48 or
3C 286) at 20 cm and 6 cm is presented for each observing date in Table
\ref{tb:cal}, and its flux density evolution with time since 1980 is
shown in Figure \ref{fig:cal}.  The data were processed in a standard
manner using the Astronomical Image Processing System (AIPS) provided
by NRAO.

\subsection{Error Estimates}
Error estimates are derived as in Weiler et al.\
\markcite{w91}(1991). That is, the rms ``map'' error, $\sigma_0$, in an
image only provides a lower limit to the total error, so that we add
quadratically an error of 5\% of the measured flux density to account for
inaccuracies in the flux density calibration and any absolute scale errors
in the primary calibrators. The the final errors, $\sigma_f$, are taken as 
\begin{equation}
\sigma_{f}^{2}\equiv (0.05 S_{0})^{2} + \sigma_{0}^{2}\ ,
\end{equation}
where $S_0$ is the observed flux density for SN 1980K and $\sigma_0$ is
the observed rms map error measured outside of any obvious regions of
emission.

\section{Results}
Since the previous publication of SN 1980K data in Weiler et
al.\ \markcite{w92}(1992), we have added three new VLA measurements at
20 cm, and two at 6 cm.

NGC 6946 was also surveyed with the VLA by Lacey et al.\
\markcite{ldg97}(1997) for compact sources on 1994 April 1 at 20 cm in
the ``A'' configuration, and on 1994 June 16 at 6 cm in the ``B''
configuration.  Although outside of their $9\arcmin\times9\arcmin$
region of interest centered on the nucleus of NGC 6946, SN 1980K was
detected in their survey at 20 cm.  Since the SN was $\sim 5.5\arcmin$
from their field center and suffered primary beam attenuation, their
measurement of flux density was corrected by 6.9\% with the standard
AIPS task PBCOR.  Their 6 cm observation suffered even more severe
attenuation at the location of SN 1980K, and only an upper limit on the
6 cm flux density could be established for the SN.

\begin{figure}
\plotfiddle{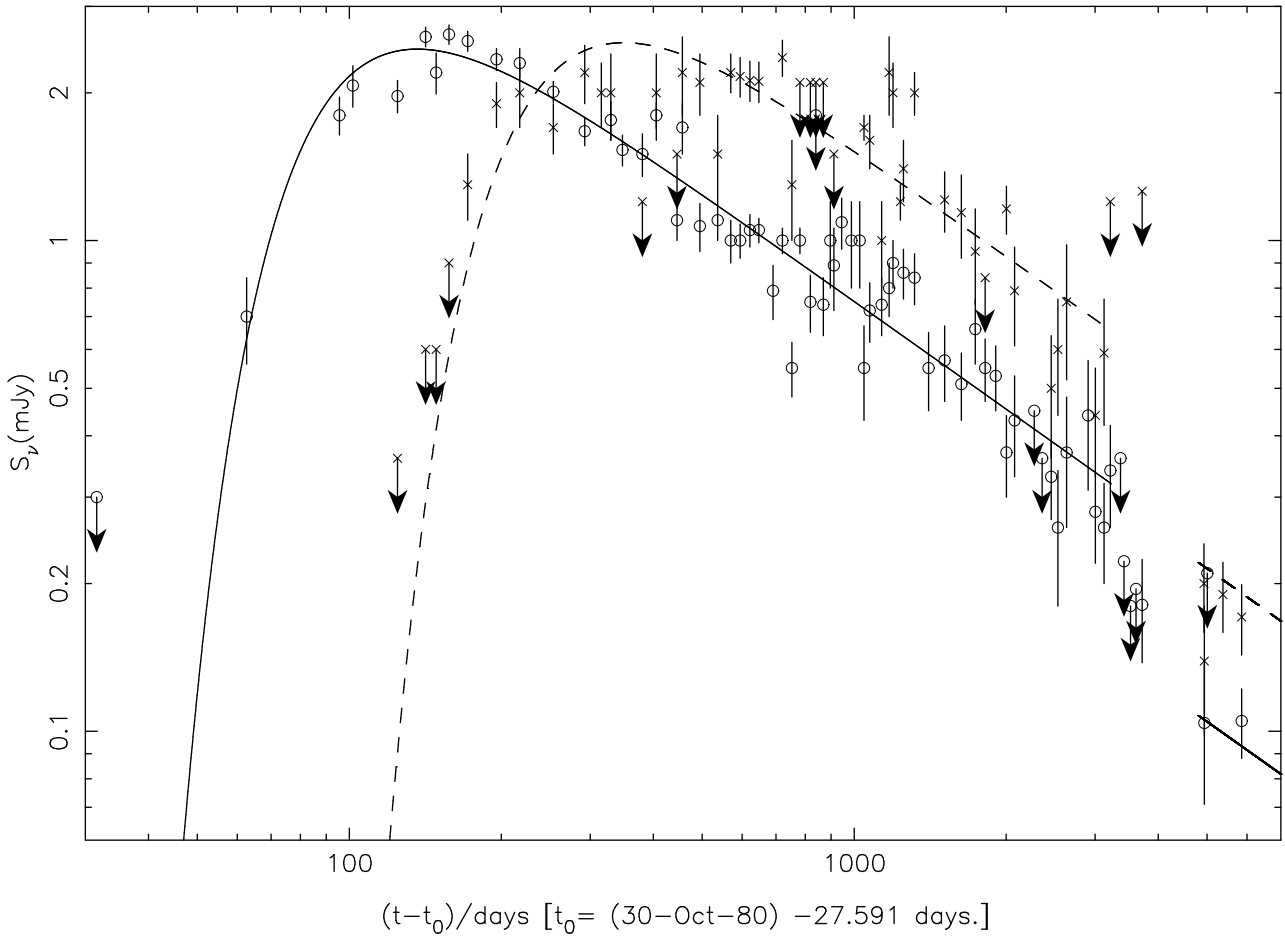}{4.5in}{270}{70}{70}{-250}{400}
\caption{
\label{fig:lc} Radio light curves for SN 1980K in
NGC 6946. The data at wavelengths 20 cm ({\em crosses}) and 6 cm ({\em
open circles}) are shown together. The data represent 16 years of
observation from 1980 November 3 through 1996 October 14, including new
and revised observations presented in this paper and  previous measurements
from Weiler et al.\ \protect\markcite{w86}(1986) and Weiler et 
al.\ \protect\markcite{w92}(1992), as well as measurements by Lacey
(\protect\markcite{l97}1997).  The {\em solid} and {\em dashed} curves
up to day 3200 represent the best-fit model light curves at 6 and 20
cm, respectively.  The curves after day 4900 are the extrapolation of
the model appropriate up to day 3200 with the same decline rate $\beta$
and same spectral index $\alpha$, but with a discontinuity of 55\%
introduced between day 3200 and day 4900.}
\end{figure}

Twelve revised measurements\footnote{\label{fn:long}We include in Table
\ref{tb:data} eight previously published measurements which were
originally listed in Weiler et al.\ (\markcite{w86}1986,
\markcite{w92}1992) as $<3\sigma$ detections. We now feel a
conservative approach to consider these data as $3\sigma$ upper limits
is more appropriate. Additionally, we have remeasured four
previously-published results and list these  revised values in Table
\ref{tb:data}. As can be seen from a comparison of Table \ref{tb:par}
in this paper with Table 4 in Weiler et al.\ \markcite{w92}(1992), the
effect of these changes on the best-fit model parameters is relatively
small.} between 1980 November 3 and 1990 December 9 and the six new
measurements plus one upper limit between 1994 April 1 and 1996 October
14 are presented in Table \ref{tb:data}. In Figure \ref{fig:lc} we plot
the entire time evolution of the flux density for SN 1980K at both 20
and 6 cm, from 1980 November 3 to 1996 October 14.  This includes the
new (and revised) results, together with earlier measurements from
Weiler et al.\ \markcite{w86}(1986) and Weiler et
al.\ \markcite{w92}(1992).  The 6 cm upper limit from Lacey
(\markcite{l97}1997) for 1994 June 16 is consistent with our
measurement on 1994 April 10.  The 20 cm measurement by Lacey for 1994
April 1 is about 1.5$\sigma$ greater than our 20 cm measurement of 1994
April 10, but the reported uncertainties overlap.

\section{Parameterized Model}

Weiler et al.\ \markcite{w86}(1986) discuss the common properties of
radio-emitting supernovae (SNe), known as radio SNe (RSNe), including
nonthermal synchrotron emission with high brightness temperature,
turn-on delay at longer wavelengths, power-law decline after maximum
with index $\beta$, and spectral index $\alpha$ asymptotically
decreasing to an optically thin value.  Weiler et
al.\ \markcite{w86}(1986) have also shown that the ``mini-shell'' model
of Chevalier (\markcite{c82a}1982a,\markcite{c82b}b) adequately
describes known Type II RSNe.  In this model, the relativistic
electrons and enhanced magnetic fields necessary for synchrotron
emission are generated by the SN shock interacting with a relatively
high-density ionized circumstellar envelope.  This dense cocoon is
presumed to have been established by a high mass-loss rate ($\dot M$~
$>$~ 10$^{-6}$ $M_\odot$~ yr$^{-1}$), low speed ($w_{\rm wind}$~
$\sim$ 10 ~km ~s$^{-1}$) wind from a red supergiant (RSG) SN progenitor
which was ionized and heated by the initial SN UV/X-ray flash.  The
rapid rise in radio flux density results from the shock overtaking
progressively more of the wind matter, leaving less of it along the
line-of-sight to absorb the emission from the shock region.

Following Weiler et al.\ \markcite{w86}(1986), we adopt the
parameterized model:
\begin{equation} 
S {\rm (mJy)} = K_1 {\left({\nu} \over {\rm 5~GHz}\right)^{\alpha}}
{\left({t - t_0} \over {\rm 1~day}\right)^{\beta}}
e^{-{\tau}} ,
\end{equation}
where
\begin{equation} 
\tau = K_2 {\left({\nu} \over {\rm 5~GHz}\right)^{-2.1}} {\left({t -
t_0} ,
\over {\rm 1~day}\right)^{\delta}} 
\end{equation}
and with $K_1$ and $K_2$ corresponding, formally, to the flux density
in mJy and the uniform external absorption, respectively,  at 5~GHz one
day after the explosion, $t_0$.  The term $e^{-{\tau}}$ describes the
attenuation of a local medium that uniformly covers the emitting source
(``uniform external absorption'').  This absorbing medium is assumed to
be purely thermal, ionized hydrogen with frequency dependence
$\nu^{-2.1}$.  The parameter $\delta$ describes the time dependence of
the optical depth for this local, uniform medium.  For an undecelerated
SN shock, $\delta = -3$ is appropriate (Chevalier
\markcite{c82a}1982a).

This parameterization has been found generally applicable to Type II
SNe, such as SN 1979C (Weiler et al.\ \markcite{w86}1986,
\markcite{w91}1991) and SN 1980K (Weiler et al.\ \markcite{w86}1986,
\markcite{w92}1992), with values of $\delta$ close to the undecelerated
value ($\delta = -3.12$ and $-2.68$ for SN 1979C and SN 1980K,
respectively; see Table \ref{tb:par}).

The best-fit model parameters are listed in Table \ref{tb:par}, only
using data through 1989 July 23. This fit is slightly different from
that of Weiler et al.\ \markcite{w92}(1992), due to improved fitting
software and reclassification and remeasurement of some data points
(see Footnote \ref{fn:long}), but the difference is not deemed to be
significant.

The errors in the fitting parameters in Table \ref{tb:par} were
estimated using a ``bootstrap'' procedure (Press et
al.\ \markcite{p92}1992). Bootstrap procedures use the actual data sets
to generate thousands of synthetic data sets that have the same number
of data points, but some fraction of the data is replaced by duplicated
original points. The fitting parameters are then estimated for these
synthetic data sets using the same algorithms that are used to
determine the parameters from the actual data. The ensemble of
parameter fits is then used to estimate errors for the parameters by
examining number distributions for the parameter in question. The
errors we present correspond to the 15.85\% and 84.15\% points in the
distribution of the occurrence of parameters from the synthetic data.

\begin{figure}
\plotfiddle{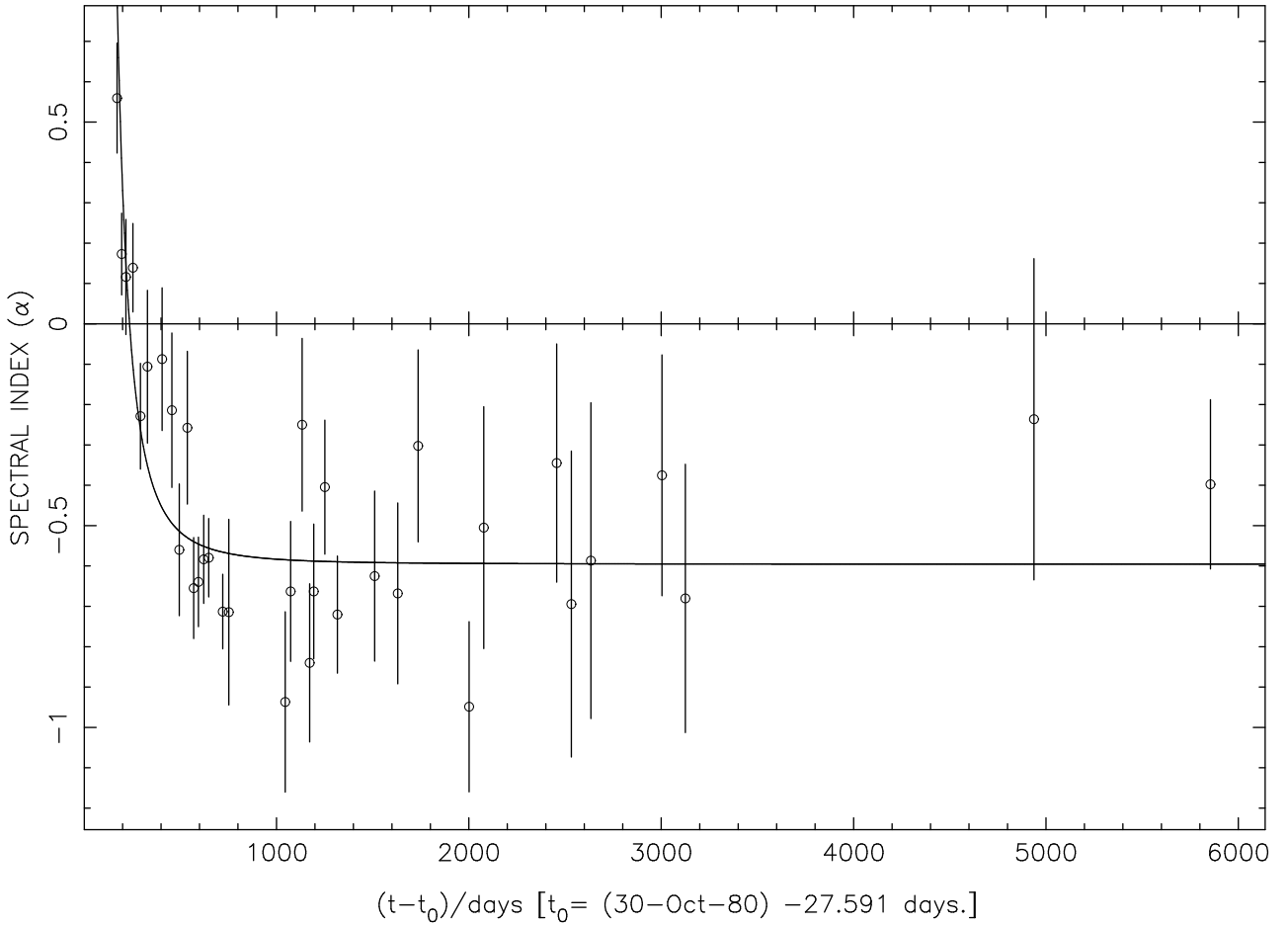}{4.5in}{270}{70}{70}{-250}{450}
\caption{
The radio spectral index, $\alpha_{6}^{20}$,
between 6 and 20 cm for SN 1980K.  Notice that, even after the the
sharp flux density decrease after day $\sim 3200$ (see
Fig.~\ref{fig:lc}), the spectral index has apparently remained constant
and is consistent, to within the errors, with the spectral index from
earlier epochs.  The {\em solid} curve is the spectral index as
calculated from the model parameters in Table \ref{tb:par}, using only
data from before ($<$ day 3200) the sharp flux density decline seen in
Fig.~\ref{fig:lc}.\label{fig:si}}
\end{figure}

Figure \ref{fig:lc} shows all of the data available through 1996
October 14, with the best-fit model up to day 3200 shown for 6 cm ({\em
solid line}) and 20 cm ({\em dashed line}).  After day 4900 we
extrapolate this model with the same decline rate $\beta$ and same
spectral index $\alpha$, but with a discontinuity of 55\% introduced
between day 3200 and day 4900.  Note that the single detection on day
3692 (1990 December 9), and several low upper limits at 6 cm, imply
that the drop-off actually started sometime between the last 6 cm
detection on day 3188 (1989 July 23) and day 3692.  Also note that the
extrapolated model light curves after day 4900, while showing the
preservation of the spectral index (see Figure \ref{fig:si}) and
consistent with the available data, are not considered to have any
predictive value, because of the large uncertainty in the data.  The
data are also consistent with a constant flux density (i.e., with
$\beta = 0$) after day 4900.

Figure \ref{fig:si} shows the evolution of the spectral
index, $\alpha_{6}^{20}$, between 6 and 20 cm; the {\em solid line} is
derived from the model fit in Table \ref{tb:par}.

\section{Discussion}

Examination of Figure \ref{fig:lc} shows that radio observations of SN
1980K, at least through day 3200, are consistent with the tenets of the
Chevalier (1982\markcite{c82a}a,\markcite{c82b}b) model, as described
by Weiler et al.\ \markcite{w92}(1992). However, sometime between day
$\sim 3200$ and day $\sim 4900$ the radio emission of SN 1980K changed
dramatically, dropping by a factor of $\sim 2$ below the values
expected from the previously best-fit model.  This roll-off is even
roughly described by the single 6 cm detection on day 3692 (1990
December 9) and the several 6 cm upper limits around that time.
Interestingly, however, the large change in flux density has apparently
not affected the spectral index evolution shown in Figure
\ref{fig:si}.  The best-fit post-decrease spectral index,
$\alpha_{6}^{20}({\rm day}>4900)=-0.42\pm 0.15$, is consistent to
within the errors with the pre-decrease spectral index,
$\alpha_{6}^{20}({\rm day}<3200)=-0.60^{+0.04}_{-0.07}$.  While such a
sharp drop could imply a change in emission mechanism, emission
efficiency, or shock transition to a non-radiative phase, we interpret
the apparently constant spectral index to imply that the basic
shock-driven emission mechanism has remained unchanged, and that the
flux density drop is most likely due to a deviation of the average
circumstellar medium (CSM) density behavior from the standard $r^{-2}$
law, expected for a pre-SN RSG wind with constant mass-loss rate
($\mdot$) and constant speed ($w$).

Since the radio luminosity of a SN is $L_{\rm radio} \propto
\left(\frac{\mdot}{w}\right)^{(\gamma-7+12 m)/4}$ ($\gamma=-2\alpha+1$
is the relativistic electron energy index and $m=-\delta/3$, where the
shock radius evolves as $R \propto t^m$, and $\delta$ is defined in \S
4; Chevalier 1982\markcite{c82a}a,\markcite{c82b}b), it is related to
the average circumstellar density ($\rho_{\rm CSM} \propto\mdot/w$), a
measure of the deviation of the radio emission from the standard model
may provide, assuming all other aspects of the emission mechanism
remain unchanged, an indication of a deviation of the circumstellar
density from the canonical $r^{-2}$ law appropriate for a constant
mass-loss rate ($\mdot$), constant speed ($w$) pre-SN stellar wind.
Thus, for SN 1980K with $L_{\rm radio} \propto (\Mdot/w)^{1.53}$, a
flux density decrease by a factor of $\sim 2$ implies, for a constant
emission mechanism, a decrease in $\rho_{\rm CSM}$ by a factor of $\sim
1.6$.  [N.B.: The expressions we use from the Chevalier model assume a
power-law CSM density distribution. They predict that a very small CS
density leads to a very small radio luminosity.  However, if there is a
sudden drop in the density, the previously accelerated electrons will
continue to radiate, and the radio luminosity will only decrease due to
adiabatic expansion. This implies that the density changes which we
obtain are, strictly speaking, lower limits, since some of the radio
luminosity must be coming from electrons that were accelerated before
the shock entered the less dense medium.]

Weiler et al.\ \markcite{w86}(1986) predicted the possibility of a
sharp decrease in the flux density for RSNe when the shock reaches the
edge of the high-density, low-speed stellar wind CSM established during
the RSG phase of the pre-SN star, and such a sharp decrease in flux
density has also been observed in SN~1957D at 6 and 20 cm (Cowan,
Roberts, \& Branch 1994\markcite{crb94}), accompanied by a steep
decline at optical wavelengths (Long, Winkler, \& Blair
1992\markcite{lwb92}).  Unfortunately, since the SN~1957D radio
measurements span only two epochs with the observations occurring
$\sim$26 and $\sim$33 years after discovery, we have no knowledge of
the prior radio behavior (spectral index evolution and temporal
history) of SN~1957D, and such a sharp drop in flux density has not
been documented in any other supernova with well-observed radio light
curves.  In at least the two cases of SN 1987A (Turtle et
al.\ \markcite{t90}1990; Staveley-Smith et al.\ \markcite{s92}1992,
\markcite{s93}1993, \markcite{s95}1995; Ball et
al.\ \markcite{b95}1995; Gaensler et al.\ \markcite{g97}1997) and SN
1979C (Montes et al.\ \markcite{m98}1998), changes in the radio flux
density evolution at ages of a few years imply that the shock is
encountering a {\em denser}, rather than less dense, CSM.

Our suggested change in the progenitor's mass-loss rate (or wind speed,
or both) would have occurred at a time before explosion of about
$t_{\rm change} \sim 11\pm 2 (v_{\rm shock}/w)$ years or, assuming
$v_{\rm shock} \sim 10^4$ km s${}^{-1}$ and $w_{\rm wind} \sim 10$ km
s${}^{-1}$, $t_{\rm change}\sim 11,000$ years before the SN.  (Note
that $m=-\delta/3 \simeq 0.9$ implies a small deceleration of the shock
with time, since the shock radius $R \propto t^m$, but given other
uncertainties in this estimate, we will not correct for that here.) The
change between implied mass-loss rates (or wind speeds, or both)
occurred within $\Delta t_{\rm change} \lesssim  4 (v_{\rm shock}/w)$
years, or $\Delta t_{\rm change}\lesssim 4000$ years with the
assumptions made above.  (On day 3200 SN 1980K still appeared to be
declining normally; by day 3700 it was already suspiciously below the
model prediction, and by day 4900 it was observed to have declined
significantly below model predictions.) The {\em minimum} time scale of
a high-to-low density transition is set by light travel time and
corresponds to $\Delta t_{\rm min} \simeq 2 R/c = 2v_{shock}/c t \simeq
0.09 t$, so that the drop over $\sim4$ years occurring $\sim11$ years
after explosion does not violate this $\Delta t_{\rm min} \gtrsim 1$
year limit.

Both the pre-1990 evolutionary phase and the 1990-to-1994
change-of-phase are short, compared to the lifetime of typical RSG
progenitor stars (5--1 Myr for stellar masses 7--15 $\Msun$,
respectively).  They are also shorter than the evolutionary time since
the last ``blue-loop'' He-burning episode (several $\times10^5$ years)
for stars with masses of $\sim$ 7--12 $\Msun$ for solar metallicity
(and higher masses for lower metallicities; see Brocato \& Castellani
\markcite{b93}1993; Langer \& Maeder \markcite{l95}1995).  The fast
drop observed for the SN~1980K radio emission, if due to a density
transition, requires a highly-structured CSM from a rapid pre-SN
stellar evolution and variations of $\Mdot/w$ by factors of at least
1.5 -- 2 over time intervals as short as 10,000 years or less.  Such
rapid phase changes have not been explored fully by stellar evolution
models, but may be similar to the well-known blue-loops occurring in
the He-burning evolutionary phase, with much shorter time scales.  It
is conceivable that these short excursions are the final phases of
evolution for massive stars, as they pass through successive nuclear
burnings (C-burning, O-burning, etc.), which are of increasingly
shorter duration and may cause the  star to quickly move in the
Hertzsprung-Russell diagram, from red to blue (or less red), and back
again.  In this context, it is interesting to note that for SN~1987A
the final blue supergiant phase, following a RSG phase, is estimated to
have lasted $<20,000$ years.  It thus appears that the final stages of
evolution of massive stars may be more complex than usually
understood.

The previously-noted change in the radio evolution of SN 1987A and SN
1979C, together with this change in SN 1980K, now provides three
examples of SNe where possible changes in the CSM density structure
have been observed and imply significant changes in stellar evolution
on time scales of only a few thousand years for the progenitors of
these SNe.  However, since the observed radio characteristics (and,
therefore, presumed CSM structure) are different in each case, there is
apparently a range of evolutionary tracks for the massive stars which
are SN progenitors.

\section{Conclusions}
At an age of $\sim 10$ years, SN 1980K has apparently, and
unexpectedly, entered a new phase of radio evolution, characterized by
a large decrease in the flux density at both 6 and 20 cm below that
expected from the model describing its previous power-law decline.  The
spectral index, however, is apparently unchanged and is consistent with
its earlier value, indicating that the emission mechanism is likely the
same.  We interpret this radio decrease to imply that the circumstellar
density has abruptly changed and the SN shock is now entering a CSM
region $\lesssim1.6$ times less dense than the RSG wind-established CSM
previously encountered by the SN shock.  This is consistent with models
for an $\sim 9 \Msun$ star that undergoes a ``blue-loop'' evolutionary
episode before entering its final RSG phase.  It is interesting to note
that changes in the radio emission, and, by implication, in the
circumstellar density structure, have also been observed on time scales
of a few years (a few $\times 10^3$ years, in the time frame of the SN
progenitor) for SN 1987A (Turtle et al.\ \markcite{t90}1990;
Staveley-Smith et al.\ \markcite{s92}1992, \markcite{s93}1993,
\markcite{s95}1995; Ball et al.\ \markcite{b95}1995; Gaensler et
al.\ \markcite{g97}1997) and SN 1979C (Montes et
al.\ \markcite{m98}1998).

Although the radio emission from SN 1980K is now very faint ($\lesssim
200\mu$Jy), we intend to continue monitoring the SN with the VLA,
to attempt to determine the reality, extent, and the radial profile for
this proposed new CSM region.  This may allow us to establish the
properties and duration of this phase of pre-SN stellar evolution, and
may, for the first time, provide a stringent constraint on the mass of
the SN progenitor based solely on the SN's radio emission.

\acknowledgements
KWW and MJM wish to thank the Office of Naval Research (ONR) for the
6.1 funding supporting this research. MJM additionally acknowledges the
support of the NRC.  SVD thanks the UCLA Astronomy Department for its
assistance, especially from Jean Turner.  We are grateful to Bruno
Leibundgut for providing the inspiration for the late-time deeper VLA
observations, to Christina Lacey for providing her VLA measurements of
SN 1980K, to Rob Fesen for a summary of recent optical measurements of
SN 1980K, and to the referee, Roger Chevalier, for his helpful comments
and suggestions on the manuscript. Additional information and data on
RSNe can be found on {\tt http://rsd-www.nrl.navy.mil/7214/weiler/} and
linked pages.

\clearpage

\clearpage

\clearpage


\clearpage
\begin{deluxetable}{llrr}
\tablewidth{0pt}
\tablecaption{Measured Flux Density Values for the Secondary Calibrator
2021+614 \label{tb:cal}}
\tablehead{
\colhead{Observation} & \colhead{Primary} & 
\multicolumn{2}{c}{Flux Density} \nl
\cline{3-4}
\colhead{Date}&\colhead{Calibrator}&
\colhead{$S_{20}$} & \colhead{$S_6$} \nl
\colhead{} & \colhead{} & \colhead{(Jy)}&\colhead{(Jy)}}
\startdata
1994 Apr 10 & 3C 286 & 2.122   & 2.936   \nl
1995 Jun 22 & 3C  48 & 2.124   & \nodata \nl
1996 Oct 13 & 3C 286 & 2.089   & \nodata \nl
1996 Oct 14 & 3C 286 & \nodata & 3.086 
\enddata
\end{deluxetable}

\clearpage
\begin{deluxetable}{lcccrr}
\tablewidth{0pt}
\tablecaption{New and Revised Flux Density Measurements for SN
1980K\tablenotemark{a} \label{tb:data}} 
\tablehead{
\colhead{Observation} & \colhead{Time Since} & \colhead{VLA} & 
\colhead{Frequency} & \multicolumn{2}{c}{Flux Density} \nl
\cline{5-6}
\colhead{Date} & \colhead{Optical Maximum\tablenotemark{b}} & 
\colhead{Configuration} &
\colhead{$\nu$}& \colhead{$S$} & \colhead{$\sigma$}\nl
\colhead{} & \colhead{(days)} & \colhead{} &
\colhead{(GHz)}& \colhead{(mJy)} & \colhead{(mJy)}}
\startdata
1980 Oct 30 & $\equiv 0$            &   &       &         &         \nl
1980 Nov 03\tablenotemark{c}   & 4    & A & 4.885   &$<$0.300  & 0.100 \nl
1981 Feb 04\tablenotemark{c}   & 97   & A & 1.465   &$<$0.360  & 0.120 \nl
1982 Nov 22\tablenotemark{c}   & 753  & D & 1.465   & $<$0.210 & 0.070 \nl
1982 Dec 30\tablenotemark{c}   & 791  & D & 1.465   & $<$0.210 & 0.070 \nl
1983 Jan 20\tablenotemark{c}   & 812  & D & 1.465   & $<$0.210 & 0.070 \nl
1983 Feb 18\tablenotemark{c}   & 841  & C & 1.465   & $<$0.210 & 0.070 \nl
1986 Dec 23\tablenotemark{c}   & 2245 & C & 4.885   &$<$0.450  & 0.150 \nl
1987 Mar 16\tablenotemark{c}   & 2328 & D & 4.885   &$<$0.360  & 0.120 \nl
1990 Feb 15\tablenotemark{d,e} & 3395 & A & 4.885   &$<$0.222  & 0.074 \nl
1990 May 29\tablenotemark{d,e} & 3498 & A & 4.885   &$<$0.180  & 0.060 \nl
1990 Aug 27\tablenotemark{d,e} & 3588 & B & 4.885   &$<$0.195  & 0.065 \nl
1990 Dec 09\tablenotemark{e}   & 3692 & C & 4.885   & 0.181    & 0.043 \nl
\tablebreak
1994 Apr 01\tablenotemark{f}   & 4901 & A & 1.450   & 0.200    & 0.041   \nl
1994 Apr 10                    & 4910 & A & 1.425   & 0.139    & 0.041   \nl
1994 Apr 10                    & 4910 & A & 4.860   & 0.104    & 0.033   \nl
1994 Jun 16\tablenotemark{d,f} & 4977 & B & 4.860   & $<$0.210 & 0.070 \nl
1995 Jun 22                    & 5348 & A & 1.425   & 0.190    & 0.031    \nl
1996 Oct 13                    & 5827 & A & 1.425   & 0.171    & 0.028    \nl
1996 Oct 14                    & 5828 & A & 4.860   & 0.105    & 0.017  
\enddata
\tablenotetext{a}{For previous measurements, see Weiler et al.\
(\markcite{w86}1986) and Weiler et al.\ (\markcite{w92}1992).}
\tablenotetext{b}{The date of the explosion is found to be 1980 October
2 (27 days before optical maximum) from the fitting process. This is
slightly different from the explosion date of 1980 October 1 assumed by
Weiler et al.\ (\markcite{w86}1986).}
\tablenotetext{c}{Measurement presented originally as a $<3\sigma$
detection which is now interpreted  as a 3$\sigma$ upper limit. See
Footnote \ref{fn:long}.}
\tablenotetext{d}{3$\sigma$ upper limit.}
\tablenotetext{e}{Data from Weiler et al.\ \markcite{w92}(1992) that
was remeasured for this paper; these values supersede those reported in
Weiler et al.\ \markcite{w92}(1992). See Footnote \ref{fn:long}.}
\tablenotetext{f}{Measurements from Lacey (\markcite{l97}1997); 
details in Lacey et al.~(\markcite{ldg97}1997).} 
\end{deluxetable}

\clearpage
\begin{deluxetable}{lcc}
\tablewidth{0pt}
\tablecaption{Fitting Parameters for SN 1980K\tablenotemark{a}\label{tb:par}}
\tablehead{
\colhead{Parameter} & \colhead{Value} & \colhead{Uncertainty
Range\tablenotemark{b}}
}
\startdata
$K_1$ (mJy)&  115               & 81--168 \nl
$\alpha$& $-0.60$              & $-$(0.56--0.67) \nl
$\beta$ & $-$0.73              & $-$(0.78--0.68) \nl
$K_2$   & $1.42\times 10^5$  & (0.60--15)$\times 10^5$ \nl
$\delta$& $-$2.68              & $-$(2.12--3.11) \nl
$t_0$   & 1980 Oct 2        & 1980 Sep 11-- Nov 1\nl
$\mdot$\tablenotemark{c}\ ($\Msun\mbox{ yr}^{-1})$ & $1.9\times 10^{-5}$ & \nodata \nl
$\chi^2/$DOF & 2.57         &\nodata  \nl
\enddata
\tablenotetext{a}{Only data through 1989 July 23 (day 3188) are used in the
fitting procedure.}
\tablenotetext{b}{The error estimates for the parameter values are
determined using the bootstrap method (see \S4).}
\tablenotetext{c}{From radio observations prior to 1989 July 23, 
employing Equation 16 from Weiler et al.\ \markcite{w86}(1986), and
assumptions of $T=2.0\times10^4\mbox{ K}, v_{\rm shock}=1.3\times
10^4\mbox{ km s}^{-1}$, and $w_{\rm wind}=10\mbox{ km s}^{-1}$.}
\end{deluxetable}

\end{document}